\documentclass[journal,twocolumn]{IEEEtran}
\usepackage{multirow}
\usepackage{mathtools, cuted}
\usepackage{amsmath}
\usepackage{bbm}
\usepackage{stfloats}
\usepackage{subfigure}
\usepackage{amsthm}
\usepackage{array,booktabs}
\usepackage{float}
\usepackage{times}
\usepackage{url}
\usepackage{amssymb,amsmath}
\usepackage{array}
\usepackage{cite}
\usepackage{dsfont}
\usepackage{multicol}
\usepackage{mathtools, cuted}
\usepackage{float}
\usepackage{breqn}
\usepackage{graphicx}
\usepackage{epstopdf}
\usepackage{mathrsfs}
\usepackage{algorithm}
\usepackage{algorithmic}
\usepackage[english]{babel}
\usepackage[autostyle]{csquotes}
\usepackage{enumerate}

\IEEEoverridecommandlockouts
\renewcommand{\algorithmiccomment}[1]{\bgroup\hfill\tiny//~#1\egroup}
\DeclareMathAlphabet\mathbfcal{OMS}{cmsy}{b}{n}

\newcommand{\argmax}{\arg\!\max}

\begin{document}
%\onecolumn
%\linenumbers
%\font\myfont=cmr12 at 20pt
\title{%\normalsize
{\huge
Learning in the Sky: An Efficient 3D Placement of UAVs}
  \vspace{-0.6cm}\hfill\\  \hfill\\
  \author{ \normalsize Atefeh Hajijamali Arani, M. Mahdi Azari, William Melek,   and Safieddin Safavi-Naeini}
% %\vspace*{-0.5 cm}
\vspace*{-0.5 cm}
}
 \IEEEaftertitletext{\vspace{-2\baselineskip}}

\maketitle

\begin{abstract}
 Deployment of unmanned aerial vehicles  (UAVs) as aerial base stations
%is considered as
%has emerged as
%provide
can deliver
a fast and
%promising
flexible
solution for
%efficiently
serving varying traffic %\textcolor{red}{demands.} \mahdi{I would use demand without s.}
demand.
%delivering high rate data services.
% However, determining an efficient 3-dimensional locations of UAVs is one of the
% main issues.
In order to adequately benefit of UAVs deployment, their efficient %three-dimensional
placement is of utmost importance, and
%Indeed, UAVs' locations
requires to intelligently adapt to the environment changes.
In this paper, we propose a
%novel and dynamic framework
learning-based mechanism
for the three-dimensional   deployment of UAVs assisting terrestrial cellular networks in the downlink. % with multiple UAVs.
The problem is modeled as a noncooperative game among UAVs in satisfaction form.
% to maximize a utility function. The utility function captures the UAV's throughput  and an activation function for the safety issue.
To solve the game,
we utilize a low complexity algorithm, in which  unsatisfied UAVs update their locations based on a learning algorithm.
%is utilized.
% For user association problem, we apply a user association metric based on the received signal power  and load.
%For user association policy, we take into account both received signal power and load.
Simulation results reveal that the proposed UAV placement algorithm
yields significant performance gains up to about
$52\%$  and $74\%$ in terms of throughput and the number of dropped users, respectively,  compared to an optimized baseline  algorithm.
%reduces average load of BSs and increases the average throughput.
\vspace*{-0.1cm}
\end{abstract}
 \begin{IEEEkeywords}
 Game theory, learning algorithm,  satisfaction, unmanned aerial vehicles.
\vspace{-0.3 cm}
 \end{IEEEkeywords}

\section{Introduction}
% The increasing demand and emerging new technologies strongly motivate the need for revolutionary approaches including development and deployment of new network architectures and  technologies. Accordingly,
Due to the novel types of services and ongoing advances in unmanned aerial vehicle (UAV) technologies, there is a growing consensus on integrating UAVs into cellular networks.
%Therefore, in
%In the next generation wireless networks,
%In this regard,
It is expected that UAVs  will
% be key components, and
play a prominent role for  traffic offloading,
capacity enhancement and disaster recovery
\cite{ 19}.
%to assist  terrestrial cellular networks in the next-generation wireless networks \cite{6}.
% for various scenarios.
% It is expected that UAVs  will be key components in the next generation wireless networks, and  play an important role  to assist  terrestrial cellular networks in various scenarios.
%UAVs are deployed for a myriad of activities such as post-disaster rescue, medical and cargo delivery, and communication assistance.
%   UAVs in the role of assisting cellular networks have numerous use cases, such as traffic offloading,
% capacity enhancement and disaster recovery
% \cite{ 10}.
From an economic perspective,    deploying small cell base stations (BSs) and/or advanced  fifth generation (5G) components, such as massive multiple-input and multiple-output (MIMO),  may not be cost-effective for temporary events. In this regard, deployment of UAVs can be considered as an
%effective and fast
alternative or complement
solution. UAVs are able to establish line-of-sight (LoS) communication links with high probability for ground users resulting in increased coverage, enhanced reliability and agility \cite{12}.
% However, there are some technical challenges
%  need to be addressed. One fundamental challenge  comes from
%  UAVs' locations which need to effectively adapt to the changes in the network’s conditions.

%  Due to the  intrinsic  characteristics  of UAVs such as flexibility and mobility,
%  %the UAV networks can be deployed and optimized
%  the  placement of UAVs can be  optimized
%  based on the  environment's conditions.
 Mobility  of UAVs  and their flexibility in adjusting their locations   significantly impact the LoS probability and the network performance.
 Most research efforts have addressed this issue from a non-learning perspective \cite{22,12,8}.
 %  to serve ground users.
 %literature, which has studied the problems related to the deployment of UAVs,
%  Existing pertinent literature can be categorized  into two classes: non-learning based approach  and  learning based approach.
% Most research efforts  have focused on non-learning based  approaches. %to  optimize the locations of UAVs.
% For instance, the authors in \cite{10} proposed a single UAV trajectory algorithm for maximizing the minimum throughput of users.  In \cite{12}, the optimum altitude of a single UAV was obtained in order  to minimize the outage probability. In \cite{8}, the authors developed an algorithm for three-dimensional (3D) placement of a UAV to maximize the number of covered users in the network. For the sake of simplicity, the problem was decoupled to the placement of the UAV in the horizontal and vertical dimensions.
However,  there has been a growing attention recently devoted  to the use of
%machine learning (ML)
learning algorithms  for the deployment problem of UAVs  \cite{4,15,18, 16}. % \cite{4,15,18,16,20}.
%for  optimizing the locations of UAVs.
%For instance,
In \cite{4} and \cite{15} learning based approaches to find the two-dimensional (2D) trajectory of UAVs flying with fixed altitudes were proposed.
% However,  optimizing  UAVs' altitudes are not considered.
%However,  they do not address optimizing  UAVs' altitudes which can impact the network performance.
%  A learning based approach for 3D placement of a single UAV is developed in \cite{17}  to enhance fairness among
% users.
 %A learning based approach for 3D placement of a single UAV is developed in \cite{18}.
%  In \cite{18} %and \atefeh{\cite{21}},
%  the authors developed learning techniques for
%  three-dimensional (3D) placement of  a single UAV  to maximize network throughput.
  In \cite{18}, a learning based approach for three-dimensional (3D) placement of a single UAV is developed  to maximize network throughput.
 % \mahdi{I remember I have sent you several papers about ML for placement of UAVs. So they do not address 3D placement? Why you have not listed them?}
%  A 3D placement algorithm for UAVs is developed in \cite{16}, in which the problem is decomposed into 2D placement and altitude selection
%It is worth mentioning that
However,
%the aforementioned
these works do not address  the 3D deployment of multiple UAVs integrated into already existing terrestrial networks.
In \cite{16}, the problem of 3D placement was dealt as a two separate optimization problems in horizontal 2D plane and altitude  of UAVs, where a prior knowledge of the   users  locations   is required. However, providing  this information   for UAVs in real time can be challenging. %and impractical.
Moreover, the proposed algorithm
is not able to adapt
%seems to be not adaptable
 to the environment changes.
% Moreover, the proposed algorithm was not designed for dynamic environments.
%which is not an online approach. %
%network deployment solution. %Furthermore, the results are limited to the cases where a prior knowledge of the network and the locations of users are available.
% \mahdi{What is your concern? Can we say: Furthermore, the other research contribution  (e.g. \cite{16}) assume that the  perfect knowledge of the network such as the locations of users is available.}
% Besides this,   they lack appropriate   channel models, in which their adopted  models do not capture the dependency of   path loss exponents  to the height of UAVs that  may have a remarkable influence \cite{1}.
Finally, to our best knowledge, the channel models in the previous
%ML
learning based reports do not capture the dependency of path loss exponents to the height of UAVs which might have a significant influence on the performance of UAVs \cite{1, 12}.

In this paper,
we address the optimal 3D locations of multiple UAVs aiming to assist the existing
%ground
terrestrial
cellular networks. %In this regard, we formulate an optimization problem that takes into account both user throughput and the traffic load. To model the problem,   we introduce a novel framework based on noncooperative game in satisfaction form which is implemented at the level of UAVs.
%To solve the game,
%In this regard,
We formulate an optimization problem using a novel framework. This framework is based on a noncooperative game in satisfaction form which is implemented at the level of UAVs.
Furthermore, as opposed to the related works, we take into account the network load effect, representing the BSs capabilities in serving users. %, in addition to the received signal power and load (BS capability) compared to classical techniques (offering the best signal strength) used in existing pertinent work.  compared to classical techniques (offering the best signal strength) used in existing pertinent work.
%To solve the game,
Then, we leverage a
%ML
learning based approach to develop a low complexity and robust algorithm allowing UAVs to
%dynamically
autonomously
adjust and optimize their locations adapted to  dynamic environments. %\atefeh{network dynamics}. %, and adapt them to the network’s conditions.
%The proposed approach offers  several distinct advantages over the aforementioned approaches. First, since only unsatisfied UAVs need to update their locations, it reduces the complexity, especially for deploying a multitude of UAVs. Second, due to
 %the fact that our proposed algorithm can be  executed in a distributed manner, the signaling overhead and exchanged information  are expected to be negligible. Furthermore, it can gain the intrinsic  advantage of distributed approaches such as improving network's robustness against failures and attacks.
 %We further elaborate on the distinct advantages of our proposed approach over the existing works.
%  \mahdi{please explain the advantages you had listed and I removed (commented), somewhere in the body of the paper. For instance after presenting the algorithm. Note that it is important to mention them.}
% Our user association policy captures both received signal power and load (BS capability) compared to classical techniques (offering the best signal strength) used in existing pertinent work.
%in order to adaptively optimize the  location of UAVs.
% Moreover, we apply a  user association policy capturing both received signal power and load.
%
% To evaluate the performance of our algorithm,
Finally, in order to examine our proposed algorithm, we employ a third generation  partnership project (3GPP)-based height dependent channel model. % similar to \cite{1}.
% which is ignored in the existing learning based pertinent literature.
Our findings from the simulation results show that the proposed learning based approach significantly improves the performance of the network, and reduces the required number of UAVs for an arbitrary target gain.

The rest of this paper is organized as follows. In Section II, we
describe the system model.  Section
III presents the proposed algorithm for the 3D deployment of  UAVs in the network.
In Section
IV, we evaluate the performance of the proposed approach.
Finally, Section V concludes the paper.

% \textit{Notations:} The regular and boldface symbols refer to scalars and matrices, respectively. For any finite set ${\mathcal A}$, the cardinality of  set ${\mathcal A}$
% is denoted by $\left|{\mathcal A}\right|$.
%  $\boldsymbol X^T$ denotes the  transpose of matrix $\boldsymbol X$.
%  The function $\mathbbm{1}_{\phi}$ denotes the indicator function which equals 1 if event $\phi$ is true and 0, otherwise.
 %\vspace{-0.1cm}
\section{System Model}
%\vspace*{-0.2cm}
In this section, we describe the system model, including network topology, channel model, and user association method.

\textit{Network Topology}:  we consider the downlink of a cellular network
consisting of a set of terrestrial BSs $\mathcal M$,
% with
and a
set of UAVs $\mathcal U$ as aerial BSs to support the users in a particular area $\mathcal R\in \mathbb{R}^2$.
% The set of users is denoted by $\mathcal K$, and the set of users associated to BS $b\in\mathcal B$ is denoted by $\mathcal K_b$, where $\mathcal B= \mathcal M \cup \mathcal U$ representing the total BSs in the network.
The set of total users and the set of users associated to BS $b\in\mathcal B$ are  denoted by  $\mathcal K$ and  $\mathcal K_b$, respectively, where $\mathcal B= \mathcal M \cup \mathcal U$ indicates the total BSs %\footnote{Note that we use term {BS} for both terrestrial BSs and UAVs.}
in the system.
We assume that all the BSs transmit over the same channel, i.e. co-channel deployment.
Let $\boldsymbol a_b(t) = (x_b(t), y_b(t), h_b(t))$  be the location of BS $b$ at time $t$, where $(x_b(t), y_b(t))$ and $h_b(t)$ are the location of BS $b$ in the horizontal dimension and its altitude at time $t$, respectively.

\textit{Radio Propagation and Signal Quality}:
we assume that the link between each user $k\in\mathcal K$ and each BS $b\in \mathcal B$ comprises  LoS and non-LoS propagation conditions. Let $\mathrm{pr}_{bk}^{\mathrm {LoS}}(t)$ denote the probability of having a LoS link between user $k$ and BS $b$ at time $t$  which is determined as follows \cite{1}:% and \cite{2}:
\begin{equation}\label{rate}
 %  \begin{align}
%\small
\footnotesize
 \mathrm{pr}_{b,k}^{\mathrm {LoS}}(t) = \prod_{n=0}^{\lfloor\frac{r_{b,k}(t) . \sqrt{\alpha . \beta}}{1000}-1 \rfloor}\Bigg[1-\exp \Bigg(-\frac{{\Big[h_b-\frac{(n+\frac{1}{2})(h_b-h_k)}{n+1}\Big]}^2}{2 \gamma^2}\Bigg) \Bigg],
  %  \end{align}
\end{equation}
 where $\alpha$, $ \beta$ and $\gamma$ represent statistical environment-dependent parameters. Here, $h_b$ and $h_k$ are the altitude of BS $b$ and the altitude of user $k$, respectively. The horizontal location of user $k$, and
 its horizontal distance to BS $b$ at time $t$ are denoted by
 $(x_k,y_k)$ and  $r_{b,k}(t)=\sqrt{(x_b(t)-x_k)^2+(y_b(t)-y_k)^2}$, respectively.  Consequently,
%  the probability
%   for the link between  BS $b$ and user  $k$ to contain a non-LoS component can be calculated as
  the non-LoS probability can be determined as
 $\mathrm{pr}_{b,k}^{\mathrm {NLoS}}(t) = 1- \mathrm{pr}_{b,k}^{\mathrm {LoS}}(t)$.
%Let the superscripts $L$ and  $N$ denote LoS and non-LoS component, respectively.

Let $d_{b,k}(t) = \sqrt{r_{b,k}^2(t) + (h_b(t) - h_k)^2}$ denote the 3D distance between BS $b$ and user $k$ at time $t$.
The path loss between BS $b$ and user $k$ can be expressed as \cite{1}:
  \begin{equation}\label{rate}
L_{b,k}^{z}(t) = A_{b}^{z} + 10 \delta_{b,k}^{z} . \log_{10}(d_{b,k}(t)),
\end{equation}
where superscript $z\in\{\mathrm{LoS}, \mathrm{NLoS}\}$ denotes LoS and non-LoS components on the link.  Here, $A_{b}^{z}$ and $\delta_{b,k}^{z}$ are the reference path loss and the path loss exponent, respectively.
Therefore,  the signal to interference
plus noise ratio (SINR)  experienced by user $k$ can be formulated as:
\begin{equation}\label{eq148}
\begin{aligned}
\gamma_{b,k}(t)=\frac{p_b(t) g_{b,k}^z(t)}{\sum_{b^\prime\in\mathcal B\backslash b} p_{b^\prime}(t) g_{b^\prime,k}^{z^\prime}(t) +\sigma^2},
\end{aligned}
\end{equation}
where $p_b(t)$ and $g_{b,k}^z(t) = 10^{-\frac{L_{b,k}^{z}(t)}{10}}$ are the transmit power of BS $b$ and the channel gain between BS $b$ and user $k$, respectively. Parameter $\sigma^2$ represents the additive white Gaussian noise (AWGN) power.

The achievable data rate provided by BS $b$ to  user $k$  using
Shannon’s capacity formula is given by:
\begin{equation}\label{eq-rate}
\begin{aligned}
R_{b,k}(t)={\omega}\log_2(1+\gamma_{b,k}(t)),
\end{aligned}
\end{equation}
where ${\omega}$ is the total bandwidth.

\textit{User Association Policy}:
  we utilize a  user association policy capturing both received signal power and load (which represents the BSs capabilities for serving new users). %compared to classical techniques offering the best signal strength. % \atefeh{used in existing pertinent work?}.
Let $\upsilon_k$ denote the traffic influx rate of user $k$. The fraction of time BS $b$ requires  to serve the traffic $\upsilon_k$  to the location
of user $k$ is defined as $\frac{\upsilon_k}{R_{b,k}(t)}$. Therefore, the load of BS
$b$ at time $t$ is given by
$\rho_b(t)=\sum_{k\in\mathcal K_b}\frac{\upsilon_{k}}{R_{b,k}(t)}$ \cite{5}.

To associate the users to the
%terrestrial BSs and the UAVs,
BSs we assume that  each BS $b$ broadcasts its estimated load, ${\hat{\rho}}_b(t)$, at time $t$ which is obtained
% through control messages. Thus, each BS $b$ computes its estimated load, ${\hat{\rho}}_b(t)$, at time $t$
as follows\cite{5}:
\begin{equation}\label{eq17}
{\hat{\rho}}_b(t)=
\eta_b(t) \rho_b(t-1)+ \Big(1-\eta_b(t)\Big)\hat{\rho}_b(t-1),
\end{equation}
where $\eta_b(t)$ is the learning rate of the load estimation for BS $b$.
Then, each user $k$ selects its serving BS based on the received signal power and the load of the BSs as follows \cite{5}:
\begin{equation}\label{ue}
\begin{split}
b_k (t) ={\mathop{\argmax }_{b\in\mathcal B}}\Big\{p_b(t) g_{b,k}^z(t) {\left(1-\hat\rho_b(t)\right)}\Big\}.
\end{split}
\end{equation}
\section{%Learning Based  Deployment of UAVs:  Towards Self-Organizing Networks
Learning Based Placement Algorithm}
In this section, we aim at optimizing the 3D locations of the UAVs for maximizing the throughput, without global network information.
%in a  distributed manner.
The problem of finding optimum 3D locations of the UAVs is  complex  mainly due to the mobility of the UAVs and temporal traffic statistics.
%We formulate the optimization problem for 3D locations of the UAVs, and solve it by leveraging machine learning.
Therefore, we leverage  the tools of
%ML
machine learning to solve the problem.
In this regard, for each UAV $u\in\mathcal U$,  a utility function is defined  as follows:
\begin{equation}\label{utility}
\begin{split}
f_u(t) = \phi_u . \frac{\sum_{k\in\mathcal K_u}R_{u,k}(t)}{\mu_u(t)} - \varphi_u . \Gamma_u(t),
\end{split}
\end{equation}
where  $\phi_u$ and $\varphi_u$ are the weight parameters that indicate, respectively, the impact of throughput  and the activation function $\Gamma_u(t)$ on the utility,  and
%In \eqref{utility},
$\mu_u(t)$ is a normalization parameter.
Here,
$\Gamma_u(t)$ is an activation function for the safety of UAVs in order to avoid collision, which is defined as follows\cite{4}:
\begin{equation}\label{activation-function}
  \Gamma_u(t) =
  \begin{cases}
  1, & d_{u,u^\prime} < d_{\mathrm{min}}, \  \ \forall u^\prime\neq u\\
  0, & \mbox{otherwise},
  \end{cases}
\end{equation}
 where $d_{u,u^\prime}$ and $d_{\mathrm{min}}$ represent the 3D distance between UAV $u$ and UAV $u^\prime$, and a certain minimum distance, respectively.
 According to \eqref{activation-function}, if the distance between UAV  $u$ and UAV $u^\prime$ is less than a minimum distance $d_{\mathrm{min}}$,  the function $\Gamma_u(t)$ returns   value one which is considered as a cost in the utility function defined in \eqref{utility}.
 %In \eqref{utility}, $\phi_u$ and $\varphi_u$ are the weight parameters that indicate the impact of throughput  and the activation function on the utility, respectively. Here, $\mu_u(t)$ is a normalization parameter.
 Let $\boldsymbol A(t) = (\boldsymbol a_1(t), \dots, \boldsymbol a_{|\mathcal U|}(t))$ denote the locations of the  UAVs at time $t$.
 Therefore, the optimization problem is formulated as follows:
  \begin{subequations} \label{max}
%\vspace*{-0.4cm}
\begin{align}
\max _{\boldsymbol A(t)}
~~ & \sum_{t\in T} \sum_{u\in\mathcal U}  f_u(t)\\
 %~~ &  \sum_{u\in\mathcal U}  f_u(t)\\
\text { s.t. } &  (x_u(t),y_u(t))\in \mathcal R, \quad \forall u \in \mathcal U, \label{eqe1}\\
& h_u\in[h_{\mathrm{min}},h_{\mathrm{max}}], \quad \forall u \in \mathcal U, \label{eqe2}\\
& 0\leq \rho_b\leq 1, \quad \forall b \in \mathcal B, \label{eqe3}
\end{align}
\end{subequations}
where $T$ is the total working time of the UAVs. The parameters $h_{\mathrm{min}}$ and $h_{\mathrm{max}}$ denote the minimum and maximum altitude of the UAVs, respectively.
For the optimization problem \eqref{max}, the constraints in \eqref{eqe1}-\eqref{eqe2} define the feasible 3D space for the locations of the UAVs.
 The constraint in \eqref{eqe3} corresponds to the definition of load, in which it avoids outages and ensures service for the users in the network.

The problem of 3D deployment of the UAVs can be formulated as a noncooperative game in satisfaction form.
In a satisfaction-form game, each UAV is interested in the satisfaction of its constraints. % instead of optimizing them.
   A game in satisfaction form can be described by the
%   following triplet \cite{5}:
% \begin{equation}\label{NFgame}
% \mathcal G =\left\langle \mathcal U, \{\mathcal S_u\}_{u\in\mathcal U}, \{c_u\}_{u\in\mathcal U} \right\rangle,
% \end{equation}
$\mathcal G =\left\langle \mathcal U, \{\mathcal S_u\}_{u\in\mathcal U}, \{c_u\}_{u\in\mathcal U} \right\rangle,$
where the set of the UAVs, $\mathcal U$, is considered as the set of players in the game. The set $\mathcal S_u$ denotes the set of strategies
%the strategy set
for UAV $u\in\mathcal U$. %, which is defined as the movement in different directions.
 We define the set $\mathcal S_u$ as the movement in different directions:
$\mathcal S_u = \{\mathrm {up, down, left, right, forward, backward, no \ change}\}$. Let  $\boldsymbol s_{-u}$ be the strategies of all UAVs except UAV $u$.
Here, the correspondence $c_u(\boldsymbol s_{-u})\subseteq \mathcal S_u$
denotes the set of strategies that can satisfy the constraints of UAV $u$  given the  strategies played by all other UAVs.
Therefore, the  correspondence $c_u(\boldsymbol s_{-u})$ can be defined as follows \cite{5}:
\begin{equation}\label{corr}
c_u(\boldsymbol s_{-u})=\{s_u\in\mathcal S_u: f_u(t)\geq\gamma_u\},
\end{equation}
where $\gamma_u$ is the satisfaction threshold for UAV $u$.
According to the observed utility, each UAV $u\in\mathcal U$ updates a satisfaction indicator $\vartheta_u(t)$ at time $t$ as follows:
\begin{equation}\label{satis-ind}
 \vartheta_u(t)=
  \begin{cases}
   1, & \mbox{if } s_u(t) \in c_u(\boldsymbol s_{-u})  \\
   0 , & \mbox{otherwise},
  \end{cases}
\end{equation}
where $s_u(t)$
%$s_u(t) \in \mathcal S_u$
is the strategy of UAV $u$  at time $t$.
For a game in satisfaction form, an  important outcome
%An important outcome of the game
is called \textit{satisfaction equilibrium} where all players are satisfied, and  $c_u(\boldsymbol s_{-u})$ is not empty for each player $u$. The notion of satisfaction equilibrium can be formulated as a fixed point as follows \cite{5}:% \cite{13}:

\textit{Definition 1 (Satisfaction Equilibrium):}  A strategy profile $\boldsymbol s^*=(s_u^*, \boldsymbol s_{-u}^*)$ is a satisfaction  equilibrium if $\forall u\in\mathcal U$, $s_u^*\in c_u(\boldsymbol s^*_{-u})$.
% \begin{equation}\label{corr1}
% s_u^*\in c_u(\boldsymbol s^*_{-u}).
% \end{equation}
%
\begin{algorithm}[t]
\scriptsize
\caption{: \footnotesize %Satisfaction
Proposed
algorithm for the 3D deployment of UAVs}
\label{alg_satis}
\begin{algorithmic}[1]
%
%\small
\STATE \textbf{Input}:  $t=0$, $\pi_{u,i}(t)$, $\vartheta_u(t-1)$, $\gamma_u$, $\forall u \in \mathcal U$ and $\forall s_{u,i} \in \mathcal S_u$.

\STATE \textbf{Output}: $\pi_{u,i}(t+1)$, $\vartheta_u(t)$, $\forall u \in \mathcal U$ and $\forall s_{u,i} \in \mathcal S_u$.\\
\WHILE{$t<T$}
\STATE $t \leftarrow t+1$
\FOR{\textbf{each} $u \in \mathcal U$}
\IF{$\vartheta_u(t-1) = 1$}
\STATE $s_u(t) = s_u(t-1)$,
\ELSE
\STATE Select a strategy $s_u(t)$
based on  $\boldsymbol\pi_u(t)$,
\ENDIF
\STATE Update the location $\boldsymbol{a}_u(t)$ based on  $s_u(t)$,
\STATE Calculate utility $f_u(t)$ according to \eqref{utility},
\STATE Calculate  $\vartheta_u(t)$ according to \eqref{satis-ind},
\STATE Update $\boldsymbol\pi_u(t)$ according to \eqref{up_pro},
\ENDFOR
\ENDWHILE
\end{algorithmic}%\vspace*{-0.2cm}
\end{algorithm}%

However, for a given satisfaction threshold, some UAVs may be faced with the situations where they can not be satisfied. Therefore, a satisfaction equilibrium may not always exist. In this context, we can reduce the satisfaction threshold $\gamma_u $ for unsatisfied UAVs after a certain time interval. %$T_{\mathrm S}$.
Therefore, we use an adaptive threshold approach described in \cite{5}.
To solve the game in a distributed manner, we use a learning algorithm, and  assume that each UAV $u$ can observe its  obtained utility. % and selected strategy.
 %For each UAV $u$,
 If the UAV is satisfied with its current utility (i.e. $\vartheta_u(t) = 1$),  the UAV has no incentive to change its location. Otherwise, it may change its location
 %will select its strategy
 according to the probability distribution
 %$\boldsymbol \pi_u(t)=(\pi_{u,1}(t), \dots, \pi_{u,|\mathcal S_u|}(t))$
 $\boldsymbol \pi_u(t)=(\pi_{u,1}(t), \dots, \pi_{u,7}(t))$, where $\pi_{u,i}(t)$ is the probability assigned to strategy  $s_{u,i}\in \mathcal S_u$ with $i\in\{1, \dots, 7\}$.
 %assigned to the set of strategies.
 Therefore, each UAV $u$ updates the probability assigned to each strategy $s_{u,i}\in \mathcal S_u$ as follows \cite{5}:
\begin{equation} \label{up_pro}
  \pi_{u,i}(t+1)= \begin{cases}
 \pi_{u,i}(t), & \mbox{if } \vartheta_u(t) = 1 \\
L_u(\pi_{u,i}(t)) & \mbox{otherwise}.
  \end{cases}
\end{equation}
Here,
\begin{equation}
{%\scriptstyle
%\displaystyle
L_u\Big(\pi_{u,i}(t)\Big)  = \pi_{u,i}(t)+\mu_u(t) q_u(t) \Big(\mathbbm{1}_{\{s_u(t)=s_{u,i}\}}-\pi_{u,i}(t)\Big),
  }
 % \vspace*{-0.1cm}
\end{equation}
where $\mu_u(t)$ is the learning rate of UAV $u$. The parameter $ q_u(t)$ is computed as  $q_u(t) = \frac{f_{\mathrm{max}} + f_u(t) - \gamma_u}{2 f_{\mathrm{max}}}$, where $f_{\mathrm{max}}$ is the maximum utility that UAV $u$ can achieve.
The function $\mathbbm{1}_{\phi}$ denotes the indicator function which equals 1 if event $\phi$ is true and 0, otherwise.

The pseudocode for the proposed   approach % for 3D deployment of the UAVs
is presented in Algorithm~\ref{alg_satis}.
%
% \begin{algorithm}[t]
% \scriptsize
% \caption{: \footnotesize %Satisfaction
% Proposed
% algorithm for 3D deployment of UAVs}
% \label{alg_satis}
% \begin{algorithmic}[1]
% %
% %\small
% \STATE \textbf{Input}:  $t=0$, $\pi_{u,i}(t)$, $\vartheta_u(t-1)$, $\gamma_u$, $\forall u \in \mathcal U$ and $\forall s_{u,i} \in \mathcal S_u$.
% \STATE \textbf{Output}: $\pi_{u,i}(t+1)$, $\vartheta_u(t)$, $\forall u \in \mathcal U$ and $\forall s_{u,i} \in \mathcal S_u$.\\
% \WHILE{$t<T$}
% \STATE $t \leftarrow t+1$
% \FOR{\textbf{each} $u \in \mathcal U$}
% \IF{$\vartheta_u(t-1) = 1$}
% \STATE $s_u(t) = s_u(t-1)$,
% \ELSE
% \STATE Select a strategy $s_u(t)$
% based on  $\boldsymbol\pi_u(t)$,
% \ENDIF
% \STATE Update the location $\boldsymbol{a}_u(t)$ based on  $s_u(t)$,
% \STATE Calculate utility $f_u(t)$ according to \eqref{utility},
% \STATE Calculate  $\vartheta_u(t)$ according to \eqref{satis-ind},
% \STATE Update $\boldsymbol\pi_u(t)$ according to \eqref{up_pro},
% \ENDFOR
% \ENDWHILE
% \end{algorithmic}%\vspace*{-0.2cm}
% \end{algorithm}
%\textit{Remark:}
The algorithm converges to an equilibrium of the game $\mathcal G$ in finite time which  is proved in ~\cite[Theorem 2]{5}.

%\atefeh{
Note that the proposed approach offers several distinct advantages. First, since only unsatisfied UAVs need to update their locations, it reduces the complexity, especially for deploying a multitude of UAVs. Second, due to the fact that our proposed algorithm can be executed in a distributed manner, the signaling overhead and exchanged information are expected to be negligible. Furthermore, it can gain the intrinsic advantage of distributed approaches such as improving network's robustness against failures and attacks.%}

%\vspace{-0.1cm}
\section{Simulation Results}
To evaluate the performance of our proposed approach, we consider a hexagonal layout  with radius $250$ m.
%, in which
The set of users are uniformly distributed in the  area, and
one terrestrial  BS located  in the center of the area.
% The transmit power of terrestrial BS and the UAVs are $46$ dBm and $24$ dBm, respectively.
The simulation parameters are
summarized in Table \ref{T2}. Furthermore, we demonstrate the performance gain of our proposed learning based scheme over the following benchmark references:  1) Strategic horizontal method proposed in \cite{11} with fixed altitude $100$ m given ${79}$ predefined potential horizontal locations,
%with $50$ m space between points,
referred to hereinafter as
\enquote{\textit{strategic}} approach. This approach is a heuristic approach, in which the horizontal location  of a new UAV is determined from  the predefined  horizontal locations to achieve the furthest distances from the BSs in the system.
% In this
% approach, the horizontal location  of a new UAV is determined based on the furthest apart
2) Random horizontal placement with  fixed altitude $100$ m, referred to hereinafter as \enquote{\textit{random-fixed altitude}} placement algorithm. 3) Random horizontal and altitude placement given ${79}$  and ${28}$ predefined potential horizontal and altitude locations,  respectively, referred to hereinafter as \enquote{\textit{random}} approach.

\begin{table}[t] \label{t1}
% \tiny
 \scriptsize	{
\begin{center}
\caption{System-Level Simulation Parameters %\cite{1,14}} \label{T2}
\cite{1,5}} \label{T2}
%\vspace*{-0.1cm}
\def\arraystretch{1.2}
\begin{tabular}{|p{0.5in}|p{1.24in}|p{1.25in}|} \hline
%\begin{tabular}{|p{0.6in}|p{1.2in}|p{1.2in}|} \hline
\multicolumn{3}{|c|}{\textbf{System Parameters}} \\ \hline \hline
\multicolumn{2}{|p{1.8in}|}{\textbf{Parameter}}& \textbf{Value}  \\  \hline
\multicolumn{2}{|p{1.8in}|}{Carrier frequency $(f_c)$/ Channel bandwidth} & $2$ GHz/ $10$ MHz \\  \hline
\multicolumn{2}{|p{1.8in}|}{Noise power spectral density} & $-174$ dBm/Hz \\ \hline
\multicolumn{2}{|p{1.8in}|}{$\upsilon_{k}$}  & $1800$ Kbps \\   \hline
\multicolumn{2}{|p{1.8in}|}{Learning rate exponent for $\eta_b(t)$} & $0.9$ \\  \hline
\multicolumn{2}{|p{1.8in}|}{$d_{\mathrm{min}}$} & $10$ m \\ \hline
\multicolumn{2}{|p{1.8in}|}{$h_{\mathrm{min}},h_{\mathrm{max}}$} & $22.5$ m, $300$ m  \\  \hline
\multicolumn{2}{|p{1.8in}|}{Altitude of users } & $1.5$ m  \\  \hline
\multicolumn{2}{|p{1.8in}|}{Altitude of  terrestrial BS} &  $25$ m  \\  \hline
\multicolumn{3}{|c|}{\textbf{BS Parameters}} \\ \hline
\textbf{Parameter} & \textbf{Terrestrial BS} & \textbf{UAV} \\ \hline
Transmit power & $46$ dBm& $24$ dBm  \\ \hline
Reference path loss  \newline (${f_c}$ in GHz)  & LoS: $28+20\log_{10}(f_c)$\newline NLoS: $13.54+20\log_{10}(f_c)$
& LoS: $30.9+20\log_{10}(f_c)$ \newline NLoS: $32.4+20\log_{10}(f_c)$ \\ \hline
Path loss exponent  & LoS: $2.2$  \newline NLoS: $3.9$ & LoS: $2.225-0.05 \log_{10}(h_u)$\newline NLoS: $4.32-0.76 \log_{10}(h_u)$ \\ \hline
\end{tabular}
\end{center}}
%\label{T2}
\vspace*{-0.4cm}
\end{table}

% \begin{figure}[tb!]
% %\hspace*{-0.5cm}
% \centering\resizebox{3.2in}{2in} {\includegraphics
% {Load_ue.eps}}\vspace*{-0.7cm}
% \caption{Average load per BS versus the number of users, given $8$ UAVs.} \label{l-u}\vspace*{-0.4cm}
% \end{figure}

% In Fig. \ref{l-u}, we depict the average load per BS versus different number of users for a network with $8$ UAVs.
% We can see that as the number of users increases, the average load increase. The learning based approach  decreases the average load compared to the other approaches. However, for the high number of users,  the behaviour of the proposed mechanism becomes closer to the benchmark algorithms, and approaches to the maximum load.

\begin{figure}[t!]
%\hspace*{-0.5cm}
\centering\resizebox{3.1in}{1.8in} {\includegraphics
{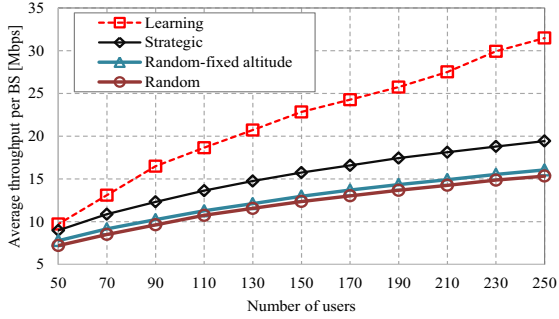}}\vspace*{-0.4cm}
\caption{Average throughput per BS versus the number of users, given $8$ UAVs.} \label{t_u}\vspace*{-0.5 cm}
\end{figure}

Fig. \ref{t_u} shows the average throughput per BS versus the number of users for a network with $8$ UAVs. The figure shows  that the learning based UAV placement approach significantly improves the average throughput compared to the benchmark algorithms. For instance, for a network with $210$  users, the proposed approach improves the
average throughput  $51.82\%$,  $84.35\%$, and $92.96\%$ compared to the strategic, random-fixed altitude, and random approaches, respectively. The main reason is that the UAVs adjust their locations in order to maximize the utility function defined in \eqref{utility} which comprises their throughput. %

\begin{figure}[tb!]
%\hspace*{-0.5cm}
\centering\resizebox{3.1in}{1.8in} {\includegraphics
{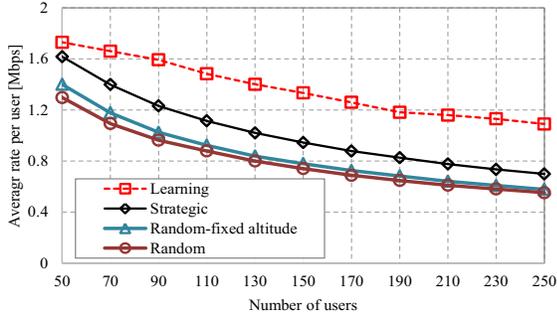}}\vspace*{-0.4cm}
\caption{Average rate per user versus the number of users, given $8$ UAVs.} \label{r-u}\vspace*{-0.6cm}
\end{figure}

Fig. \ref{r-u} illustrates the average rate per user.
%versus the number of users for  given $8$ UAVs.
We can observe that as the number of users increases, average rate per user decreases due to increasing load and the availability of limited resource in the network.
% Furthermore, the learning based UAV placement algorithm improves the average rate of users compared to the benchmark algorithms.
Since the learning based approach optimize the locations of UAVs in terms of maximizing throughput,  it improves the average rate of users compared to the benchmark algorithms
% For instance, for  a network with $190$ users, the proposed approach yields $43.05\%$, $73.04\%$, and $82.51\%$ average user's rate improvement compared to strategic, random-fixed altitude, and random approaches, respectively. %The improvement in user's rate is achieved by  the positioning of the UAVs, in which it leads to a higher level of SINR.


\begin{figure}[t]
%\vspace*{-0.1cm}
\centering\resizebox{3.1in}{1.8in} {\includegraphics
{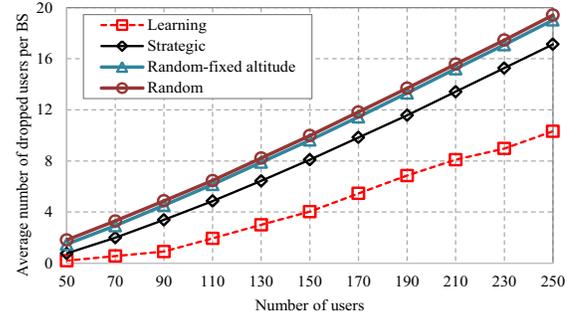}}\vspace*{-0.4cm}
\caption{Average number of dropped users per BS versus the number of users, given $8$ UAVs.} \label{drop-u}
\vspace{-0.5cm}
\end{figure}

In Fig. \ref{drop-u}, we show  the average number of dropped users per BS. %(i.e. the users not being able to served able to connect to their BS in outages due to the overloading of their serving BSs)
%resulted from overloading.
%versus the number of users.
%for a network with $8$ UAVs.
%As observed for all the approaches,
The figure shows that %as the number of users increases,
the average number of dropped users increases with an increase in the number of users.
This is mainly due to the fact that,
a limited resource is available in the network. Therefore, with increasing the number of users, some of them  may experience  reductions in their rates   due to the overloaded  BSs.
Since the proposed approach improves the  average  rate compared to the other approaches, it  yields  better performance in terms of average number of dropped users.
The performance gain of  the learning based approach compared to the strategic, random-fixed altitude, and random approaches is up to $74.14\%$, $86.76\%$, and $89.23\%$, respectively.
Accordingly, the learning approach is more resistant to the higher traffic demand.

\begin{figure}[t]
%\hspace*{-0.5cm}
\centering\resizebox{3.1in}{1.8in} {\includegraphics
{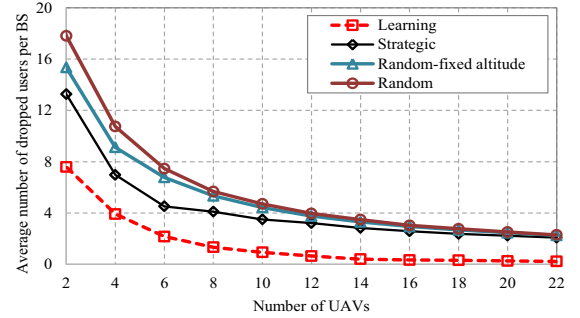}}\vspace*{-0.4cm}
\caption{Average number of dropped users per BS versus the number of UAVs, given $100$ users.} \label{drop-uav}
\vspace*{-0.5cm}
\end{figure}

Fig. \ref{drop-uav} depicts the average number of dropped users as the
number of UAVs varies. From the figure, as the number of UAVs increases the average number of dropped users per BS decreases. The main reason  is that the users associated to the highly loaded BSs can be offloaded to the lightly loaded BSs.
% Furthermore, the learning based approach yields a better performance in terms of number of dropped users compared to the benchmark algorithms.
%For instance, for a network with $4$ UAVs, the reductions of average number of dropped users in the learning based approach compared to the strategic, random-fixed altitude, and random approaches are $43.78\%$, $57\%$, and $63.48\%$, respectively.
Moreover, Fig.  \ref{drop-uav} reveals that by using the learning based approach, the number of UAVs required to reduce the number of outage users below a certain threshold is significantly lower. For instance, using learning based method only $6$ UAVs are required for less than $2$  users in outage, which is notably lower than $22$ UAVs needed in the other methods.
\vspace{-0.2cm}
\section{Conclusion}
In this paper, we have proposed a low-complexity robust algorithm for the 3D placement of UAVs integrated into  terrestrial cellular networks.
The proposed approach  leverages  the tools from game theory and
%ML
machine learning to learn optimal locations of UAVs in a distributed manner.
%Our channel model   comprises a realistic height dependent   radio propagation model. \textcolor{red}{In addition?}, the user association mechanism incorporates both the quality of the communication link and traffic load.
Our results have shown that
the  proposed  approach significantly
outperforms the benchmark algorithms in terms of both throughput and dropped users.
% Moreover, the required number of UAVs for serving ground users has been considerably lower compared to the other approaches.
We have also shown that the number of UAVs for serving ground users using the learning based approach is significantly lower than the benchmark algorithms resulting in a more cost-effective networking with UAVs.
% Furthermore,  our proposed approach does not impose any signaling overhead, and can be executed in a distributed manner.

\vspace{-0.3cm}
\bibliographystyle{IEEEtran}
\bibliography{Bib_Atefeh}

\begin{thebibliography}{10}
\providecommand{\url}[1]{#1}
\csname url@samestyle\endcsname
\providecommand{\newblock}{\relax}
\providecommand{\bibinfo}[2]{#2}
\providecommand{\BIBentrySTDinterwordspacing}{\spaceskip=0pt\relax}
\providecommand{\BIBentryALTinterwordstretchfactor}{4}
\providecommand{\BIBentryALTinterwordspacing}{\spaceskip=\fontdimen2\font plus
\BIBentryALTinterwordstretchfactor\fontdimen3\font minus
  \fontdimen4\font\relax}
\providecommand{\BIBforeignlanguage}[2]{{%
\expandafter\ifx\csname l@#1\endcsname\relax
\typeout{** WARNING: IEEEtran.bst: No hyphenation pattern has been}%
\typeout{** loaded for the language `#1'. Using the pattern for}%
\typeout{** the default language instead.}%
\else
\language=\csname l@#1\endcsname
\fi
#2}}
\providecommand{\BIBdecl}{\relax}
\BIBdecl

\bibitem{19}
M.~{Mozaffari}, W.~{Saad}, M.~{Bennis}, Y.~{Nam}, and M.~{Debbah}, ``A tutorial
  on {UAVs} for wireless networks: Applications, challenges, and open
  problems,'' \emph{IEEE Commun. Surveys Tuts.}, vol.~21, no.~3, pp.
  2334--2360, 2019.

\bibitem{12}
M.~M. {Azari}, F.~{Rosas}, K.~{Chen}, and S.~{Pollin}, ``Ultra reliable {UAV}
  communication using altitude and cooperation diversity,'' \emph{IEEE Trans.
  Commun.}, vol.~66, no.~1, pp. 330--344, 2018.

\bibitem{22}
Z.~{Wang}, L.~{Duan}, and R.~{Zhang}, ``Adaptive deployment for {UAV}-aided
  communication networks,'' \emph{IEEE Trans. Wireless Commun.}, vol.~18,
  no.~9, pp. 4531--4543, Sep. 2019.

\bibitem{8}
M.~{Alzenad}, A.~{El-Keyi}, F.~{Lagum}, and H.~{Yanikomeroglu}, ``{3-D}
  placement of an unmanned aerial vehicle base station {(UAV-BS)} for
  energy-efficient maximal coverage,'' \emph{IEEE Wireless Commun. Lett},
  vol.~6, no.~4, pp. 434--437, 2017.

\bibitem{4}
B.~{Khamidehi} and E.~S. {Sousa}, ``Reinforcement learning-based trajectory
  design for the aerial base stations,'' in \emph{2019 IEEE 30th Annual
  International Symposium on PIMRC}, 2019, pp. 1--6.

\bibitem{15}
H.~{Dai}, H.~{Zhang}, M.~{Hua}, C.~{Li}, Y.~{Huang}, and B.~{Wang}, ``How to
  deploy multiple {UAVs} for providing communication service in an unknown
  region?'' \emph{IEEE Wireless Commun. Lett.}, vol.~8, no.~4, pp. 1276--1279,
  Aug 2019.

\bibitem{18}
R.~{Ghanavi}, E.~{Kalantari}, M.~{Sabbaghian}, H.~{Yanikomeroglu}, and
  A.~{Yongacoglu}, ``Efficient {3D} aerial base station placement considering
  users mobility by reinforcement learning,'' in \emph{2018 IEEE WCNC}, April
  2018, pp. 1--6.

\bibitem{16}
H.~{El Hammouti}, M.~{Benjillali}, B.~{Shihada}, and M.~{Alouini},
  ``Learn-as-you-fly: A distributed algorithm for joint {3D} placement and user
  association in multi-{UAVs} networks,'' \emph{IEEE Trans. Wireless Commun.},
  vol.~18, no.~12, pp. 5831--5844, Dec 2019.

\bibitem{1}
M.~M. Azari, G.~Geraci, A.~Garcia-Rodriguez, and S.~Pollin, ``{UAV}-to-{UAV}
  communications in cellular networks,'' \emph{arXiv preprint
  arXiv:1912.07534}, 2019.

\bibitem{5}
A.~{Hajijamali Arani}, M.~J. {Omidi}, A.~{Mehbodniya}, and F.~{Adachi},
  ``Minimizing base stations' {ON/OFF} switchings in self-organizing
  heterogeneous networks: A distributed satisfactory framework,'' \emph{IEEE
  Access}, vol.~5, pp. 26\,267--26\,278, 2017.

\bibitem{11}
F.~{Lagum}, I.~{Bor-Yaliniz}, and H.~{Yanikomeroglu}, ``Strategic densification
  with {UAV-BSs} in cellular networks,'' \emph{IEEE Wireless Commun. Lett},
  vol.~7, no.~3, pp. 384--387, 2018.

\end{thebibliography}

\end{document}